\documentstyle{article}

\begin{document}
\title{Correlations between the interacting fragments of decaying processes}
\author{Pedro Sancho \\ GPV de Valladolid \\ Centro Zonal en Castilla y Le\'on  \\ Ori\'on 1, 47014, Valladolid, Spain}
\date{}
\maketitle
\begin{abstract}
We study the correlations (and alignment as a particular case)
existent between the fragments originated in a decaying process when
the daughter particles interact. The interaction between the particles is modeled using the potential of coupled oscillators,
which can be treated analytically. This approach can be considered
as a first step towards the characterization of realistic
interacting decaying systems, an archetypal process in physics. The
results presented here also suggest the possibility of manipulating
correlations using external fields, a technique that could be useful
to provide sources of entangled massive particles.
\end{abstract}
\vspace{7mm}

\section{Introduction}
In classical mechanics two particles originated in a decaying
process maintain perfect correlations between their positions and
momenta during the subsequent evolution of the system. In quantum
mechanics the situation is not so favorable. The spread of the
wavefunction and the uncertainties in the initial positions and
momenta can cause in many cases a rapid lost of the initial
correlations \cite{per,Per,bel}. In these cases the correlations are
not useful for practical purposes. References \cite{Per,bel}
studied free non-relativistic decaying systems. Real systems are not in general free,
but present various types of interactions. We pose in this paper the
question if, when these interactions are taken into account, the
picture of the process and the behavior of the correlations, are
modified. The study of real interactions is in general very
complicated, and many times we must resort to perturbative methods.
We propose here an alternative approach. We consider a simple model
of interaction that can be solved analytically by assuming that the
two particles interact as coupled oscillators. In all the areas of
physics, but in particular in quantum mechanics, many problems that
are not analytically treatable are reduced to harmonic oscillators.
As we are dealing with two particles, coupled oscillators could be a
good candidate for our problem. On the other hand, coupled
oscillators have been used in several branches of physics to
simulate more complex systems \cite{Iac, Han1, Han2}. We remark that
we are modeling the evolution taking place after the decaying
process, not the transformation of the mother into two daughter
particles, a non-unitary process incompatible with the unitary
evolution of coupled oscillators. The model of interaction not only
holds for the direct interaction between the two particles (the
electromagnetic interaction in the case of charged particles) but
also for external fields, which could be introduced with the aim of
driving the evolution of the system (see, for instance, Ref.
\cite{PRL} where external fields are used to prepare non-spreading
wave packets).

In addition to the more complete characterization of decaying
systems (one prototype example in physics) provided by the inclusion
of interactions, the analysis below allows for a better
understanding of how correlations could be manipulated. In
connection with the last point and from a more practical point of
view, it must be remarked that decaying systems with good
correlations could provide a source of entangled massive particles,
well suited for two-particle interference experiments. Up to now,
most two-particle interference arrangements use entangled photons
because of the difficulty in preparing massive entangled states.

The plan of the paper is as follows. First, in Sect. 2, we briefly
review the basics of decaying systems. The model of interaction,
based on coupled oscillators, is presented in Sect. 3. In Sect. 4 we
derive the states of minimum uncertainty, which reach the best
compromise between momentum and position initial uncertainties. Our
analysis is restricted to these states. The evolution of the
interacting states is evaluated in Sect. 5. Section 6 deals with the
comparison between perfect correlations and alignment for
interacting and free systems. In Sect. 7 we evaluate the
entanglement present in the system using the Schmidt number.
Finally, in the Conclusions we discuss the principal results of the
paper.

\section{Basics of decaying systems}

We present in this Section a brief review of the problem. The
classical variables describing a decaying process are
${\bf x} + {\bf y}$ and ${\bf p_{x}}+{\bf p_{y}}$, with ${\bf x}$ and ${\bf y}$ the positions of the particles and ${\bf p_{x}}$ and ${\bf p_{y}}$
their momenta. In classical theory the law of momentum
conservation implies that if initially the total momentum
is ${\bf p_{x}}+{\bf p_{y}}={\bf 0}$, it will remain so during
the subsequent free evolution of the complete system; as ${\bf
p_{x}}=-{\bf p_{y}}$ the particles will always be found in
opposite directions.

In quantum theory ${\bf x} + {\bf y}$ and ${\bf p_{x}}+{\bf
p_{y}}$ are replaced by the operators $\hat{{\bf x}} + \hat{{\bf
y}}$ and ${\bf \hat{p}_{x}}+{\bf \hat{p}_{y}}$. The uncertainty
relations for these operators are
\begin{equation}
\Delta (\hat{x}_i + \hat{y}_i ) \Delta (\hat{p} _{x_i} + \hat{p}
_{y_i}) \geq \hbar \label{eq:ww}
\end{equation}
with $i=1,2,3$ the three components of each operator and $\Delta
\hat{x} _i =<(\hat{x}_i -< \hat{x}_i >  )^2 >^{1/2} ,\cdots $. Because
of these relations it is not possible to prepare the system in a
state with perfect correlations in momentum, as it was the case in
classical theory ($ \Delta (\hat{x} _i + \hat{y}_i)$ would have an
unbounded value). The most one can expect in realistic conditions is
to have a sharp distribution centered around ${\bf p_{x}}+{\bf
p_{y}}={\bf 0}$. For instance, in Ref. \cite{Per} the wavefunction
of the system is taken as $\psi ({\bf x},{\bf y},t)= \int \int
f({\bf p_x},{\bf p_y}) \exp{(i({\bf p_x}\cdot {\bf x}+{\bf p_y}\cdot {\bf
y}-Et)/\hbar )}d^3{\bf p_x} d^3{\bf p_y}$ with $E$ the energy of the
system and $f$ the momenta distribution, sharply centered around
${\bf p_{x}}+{\bf p_{y}}={\bf 0}$.

As a consequence of this uncertainty in the initial momenta and the
inherent spreading of the wavefunction the correlations present in
the classical case can be lost in many situations in the quantum
realm. We discuss this point at extent in Sect. 6.

\section{Interaction: coupled oscillators}

Up to our knowledge in all the studies of decaying systems presented
so far in the literature the subsequent evolution of the system is
assumed to be free. However, in realistic conditions, in most of the
cases there is some type of interaction between the products of the
process. For instance, if a neutral particle decays into two
particles with opposite charges there is an electromagnetic
interaction which, at least at the initial times, is not negligible.
The exact description of the system taking into account the
electromagnetic interaction is, however, too complex.

We propose here a different approach. Instead of studying particular
forms of realistic interactions we shall analyze an interaction that
can be treated analytically. This approach can be considered as a
first approximation to the problem. As signaled in the introduction
harmonic oscillators are extensively used in classical and quantum
theory to study complex systems not analytically solvable. Then we
model the inter-particle interaction via the coupled oscillator
potential.

There are other more physical reasons for this choice. Imagine we
try to introduce some external interaction, for instance through the
presence of some external fields, aimed to preserve the initial
correlations. Examples can be found in the literature of the use of
external fields to manipulate the form of the wavefunction. For
instance, an imaginary potential can compensate the spreading of the
wavefunction resulting in a non-spreading wavepacket \cite{PRL}.
Good correlations are obtained when ${\bf x}_{mea} \approx -{\bf
y}_{mea}$, i.e., when the measured positions of the particles are
close to the perfect (classical) correlation ${\bf x}_{mea} = -{\bf
y}_{mea}$. Then if we introduce a potential of the type
\begin{equation}
V({\bf x},{\bf y})=\kappa ({\bf x}+{\bf y})^2
\label{eq:UNO}
\end{equation}
we have an interaction that acts when the particles tend to
deviate from the good correlations. The strength of the
interaction is proportional to the magnitude of the deviation,
representing the constant $\kappa $ the value of this
proportionality. This is the ideal form of an interaction aimed to
maintain good correlations. It can be seen as the ideal limit
towards which any realistic interaction with this aim must tend. Other acceptable
choices should have the form $\kappa ({\bf x}+{\bf y})^{2n}$, with
$n$ any positive integer number. However, we shall restrict our
considerations to the case $n=1$.

The form of Eq. (\ref{eq:UNO}) is that of the interaction term of
two coupled oscillators. Classically, the Hamiltonian of two
coupled oscillators is:
\begin{equation}
H=\frac{{\bf p}_1^2}{2m_1} + \frac{{\bf p}_2^2}{2m_2}+C_1 {\bf x}_1^2 + C_2 {\bf x}_2^2 +C_{12}{\bf x} \cdot {\bf y}
\end{equation}
The quantum Hamiltonian is obtained from the classical one by the
usual quantization procedure. Our case corresponds to the choice
$C_1=C_2=\kappa$ and $C_{12}=2\kappa $. As signaled in the
introduction coupled oscillators have been used to simulate other
physical systems \cite{Iac, Han1, Han2}.

The choice of a coupled oscillator to simulate the interaction of
the decaying particles could be criticized because oscillators are
usually associated with states of systems that remain bound, whereas
decaying systems become free. However, it must be noted that the
oscillator-type {\it force} only acts on the center of mass
coordinate ${\bf x}+{\bf y}$, remaining the relative position
coordinate ${\bf x}-{\bf y}$ free. This way we can have a coupled
oscillator interaction and, at the same time, {\it free evolution}.
The particles can become well separated as can easily be tested from
the solutions of the problem (see comment after Eq. (\ref{eq:kk})).

\section{States of minimum uncertainty}

As remarked in Refs. \cite{Per} and \cite{bel} and previous sections the initial
uncertainty in position and momentum is one of the causes of
alignment lost in decaying systems. This uncertainty can be
minimized choosing the states of minimum uncertainty of the relevant
operators of the problem, $\hat{{\bf x}} + \hat{{\bf y}}$ and ${\bf
\hat{p}_{x}}+{\bf \hat{p}_{y}}$. These states can be easily
calculated using well-known techniques \cite{Gal}: given two
operators $\hat{Q} _+ $ and $\hat{Q} _- $ they are determined by the
equations,
\begin{equation}
(\hat{Q }_+ - Q _+  ) | \psi > =-i \epsilon  (\hat{Q
}_- - Q _-  ) | \psi >
\label{eq:cua}
\end{equation}
with
\begin{equation}
Q _{\pm} = < \hat{Q } _{\pm} >_{\psi }  \; ; \;   \epsilon  = -
\frac{<\psi |i[\hat{Q }_ + , \hat{Q }_ - ] |\psi >}{2 (\Delta _{\psi}
\hat{Q }_ - )^2}
\label{eq:cin}
\end{equation}
where $[\hat{Q }_ + , \hat{Q }_ - ]$ is the commutator of $\hat{Q }_
+ $ and $\hat{Q }_ - $ and $\Delta _{\psi} \hat{Q }_ - =<(\hat{Q}_-
-< \hat{Q}_i >  )^2 >^{1/2}$ is evaluated in the state $\psi $. In
the position representation the operators can be written as $(\hat{Q
}_+ )_j=\hat{x} _j + \hat{y}_j=x_j + y_j$ and $(\hat{Q }_-
)_j=(\hat{p} _{x} )_j + (\hat{p} _y)_j= -i\hbar
\partial /\partial x_j -i\hbar \partial /\partial y_j $, with
$j=1,2,3$. Then Eqs. (\ref{eq:cua}) and (\ref{eq:cin}) become:
\begin{equation}
(x_j +y_j - (Q _+)_j -i (Q _-)_j \epsilon _j ) \psi
= - \epsilon _j \hbar \left( \frac{\partial
\psi}{\partial x_j} +\frac{\partial \psi}{\partial y_j}  \right)
\end{equation}
and
\begin{equation}
\epsilon _j = -\frac{<\psi |-2\hbar |\psi >}{2 (\Delta _{\psi}
(\hat{p}_{x_j}+\hat{p}_{y_j} ))^2} = \frac{\hbar}{2 (\Delta _{\psi}
(\hat{p}_{x_j}+\hat{p}_{y_j} ))^2}
\end{equation}
where, and from now on, we assume by simplicity $\epsilon _1
=\epsilon _2 = \epsilon _3 = \epsilon $.

In the particular case $Q_+=Q_-=0$, it is immediate to verify by direct substitution that the solution
of the above equation is
\begin{equation}
\psi ({\bf x}, {\bf y}) \sim \exp{(-({\bf x}+ {\bf y} )^2/4
\epsilon \hbar ) }
\label{eq:once}
\end{equation}
We note that this is an entangled wavefunction, which reflects a
good initial correlation for position measurements (${\bf x}_{mea}
\approx - {\bf y}_{mea}$) with a dispersion of the order $2 \hbar
^{1/2} \epsilon ^{1/2}$.

\section{Evolution of the system}

The evolution of the state is ruled by Schr\"{o}dinger's equation
(we assume by simplicity the mass of the two particles to be the
same):
\begin{equation}
i\hbar \frac{\partial }{\partial t} \psi ({\bf x}, {\bf y},t)
=-\frac{\hbar ^2}{2m} (\nabla _{\bf x}^2 + \nabla _{\bf y}^2 )
\psi ({\bf x}, {\bf y},t) + \kappa ({\bf x}+ {\bf y})^2 \psi ({\bf
x}, {\bf y},t)
\end{equation}
In order to solve this equation it is convenient to introduce the
center of mass and relative position coordinates, defined by the
relations:
\begin{equation}
{\bf X} =\frac{1}{2} ({\bf x}+ {\bf y}) \; ; \; {\bf Y}= {\bf x}- {\bf y}
\end{equation}
The Schr\"{o}dinger equation becomes:
\begin{equation}
i\hbar \frac{\partial }{\partial t} \psi ({\bf X}, {\bf Y},t) =
-\left( \frac{\hbar ^2}{2M} \nabla _{\bf X}^2 + \frac{\hbar ^2}{2\mu
} \nabla _{\bf Y}^2  \right) \psi ({\bf X}, {\bf Y},t) +
\frac{M\omega ^2}{2} {\bf X}^2 \psi ({\bf X}, {\bf Y},t)
\label{eq:dd}
\end{equation}
where $M=2m$ is the total mass, $\mu =m/2$ is the reduced mass and
$\omega ^2 =8\kappa /M$ is a habitual way of expressing the
coefficient of the interaction potential.

The solution of this equation at time $t$ can be obtained by
integration of the initial wavefunction using the kernel or
propagator of the system \cite{Fey}:
\begin{equation}
\psi ({\bf X}, {\bf Y},t) = \int d^3 {\bf X _o} \int d^3 {\bf Y _o} \,
K({\bf X}, {\bf Y},t; {\bf X_o}, {\bf Y_o},t_o ) \psi _o({\bf X_o},
{\bf Y_o},t_o)
\end{equation}

with $\psi _o({\bf X_o}, {\bf Y_o},t_o)=N_o \exp{(-{\bf X_o}^2/a^2)}$
the initial wavefunction, given by the minimum uncertainty state
studied in Sect. 4. $N_o$ represents the normalization of this
initial wavefunction (see later). All the integrations are carried
out between the limits $-\infty $ and $\infty $. As Eq.
(\ref{eq:dd}) can be separated for variables ${\bf X}$ and ${\bf
Y}$, the kernel is the product of the kernels corresponding to the
two spatial variables \cite{Fey}:
\begin{equation}
K({\bf X}, {\bf Y},t; {\bf X_o}, {\bf Y_o},t_o ) =K_{\bf X}({\bf
X},t; {\bf X_o},t_o) K_{\bf Y} ({\bf Y},t; {\bf Y_o},t_o )
\end{equation}
Now, we have two well-known kernels. From now on we take $t_0=0$ by simplicity. For variable ${\bf X}$ it
corresponds to the kernel of the harmonic oscillator \cite{Fey}:
\begin{equation}
K_{\bf X}({\bf X},t; {\bf X_o},t_o) = \left( \frac{M\omega }{2\pi i
\hbar \sin \omega t}  \right) ^{3/2} \exp \left( \frac{iM\omega }{2
\hbar \sin \omega t} (({\bf X} ^2 +{\bf X_o}^2)\cos \omega t -2{\bf
X} \cdot {\bf X_o}) \right)
\end{equation}
On the other hand, for variable ${\bf Y}$ we have a free-particle
evolution equation with kernel \cite{Fey}:
\begin{equation}
K_{\bf Y}({\bf Y},t; {\bf Y_o},t_o) = \left( \frac{\mu }{2\pi i
\hbar t}  \right) ^{3/2} \exp \left( \frac{i\mu }{2 \hbar t} ({\bf Y}
-{\bf Y_o})^2 \right)
\end{equation}
After a simple integration using the well-known relation $\int dz
\exp (\alpha z^2 + \beta z)=(\pi /-\alpha )^{1/2} \exp (-\beta ^2
/4\alpha )$ valid for $Re(\alpha )\leq 0$, we obtain
\begin{equation}
\psi ({\bf X}, t) = N \exp{(if({\bf X}, t))} \left(
\frac{M\pi \omega }{2\pi i \hbar a^{-2}\sin \omega t + M\pi \omega
\cos \omega t}  \right) ^{3/2} \exp{(-\alpha (t) {\bf X}^2)}
\label{eq:jj}
\end{equation}
where
\begin{equation}
f({\bf X}, t) = \left( - \frac{M^3 \omega ^3}{8\hbar ^3 \sin ^3
\omega t}\left( \frac{\cos \omega t}{\frac{1}{a^4} + \frac{M^2 \omega
^2 \cos ^2 \omega t}{4 \hbar ^2 \sin ^2 \omega t}} \right) +
\frac{M\omega \cos \omega t}{2\hbar \sin \omega t} \right) {\bf X}^2
\end{equation}
and
\begin{equation}
\alpha (t)= \frac{M^2 \omega ^2 a^2}{4 \hbar ^2 \sin ^2 \omega t +
M^2 a^4 \omega ^2 \cos ^2 \omega t}
\end{equation}
$N$ is the normalization factor. On the other hand, $f$ includes all
the real functions that appear in the form $\exp (if)$ in the
wavefunction. They are of no interest because when calculating
probabilities (the magnitudes of interest in next section) they only
contribute as a constant term. Note that the wavefunction does not
depend on the ${\bf Y}$ variable. The initial wavefunction $\psi _o$
does not depend on ${\bf Y_o}$. Then as $K_{\bf Y}$ is a free
propagator cannot generate a dependence on ${\bf Y}$. In physical
terms, if the initial state does not depend on the relative
coordinates, the interaction term that is only function of the
center of mass variables cannot introduce that type of dependence
during the subsequent evolution.

The normalization factor is usually evaluated from  the condition
for the probability $\int d^3 {\bf x} \int d^3 {\bf y} |\psi ({\bf
x}, {\bf y},t )|^2 =1$. However, wavefunction (\ref{eq:jj}) cannot
be normalized in an absolute sense. Effectively, a simple
calculation gives
\begin{equation}
\int d^3 {\bf x} \int d^3 {\bf y} |\psi ({\bf x}, {\bf y},t )|^2
= |N|^2 |\mu (t)|^2 \left( \frac{\pi }{2\alpha } \right) ^{3/2} \int d^3 {\bf x}
\label{eq:ggg}
\end{equation}
where $\mu (t)$ is the term in Eq. (\ref{eq:jj}) going as the $3/2$ power of the expression between parentheses.

Expression (\ref{eq:ggg}) is unbounded because the integration is
between $-\infty $ and $\infty $. Then we must work with relative
probability densities or probability densities per unit volume,
which are given by $\int d^3 {\bf x} \int d^3 {\bf y} |\psi ({\bf
x}, {\bf y},t )|^2 / \int d^3 {\bf x}$. It is simple to see that
this relative probability is normalized to unity. The normalized (in
this relative sense) wavefunction is:
\begin{equation}
\psi ({\bf X}, t)= \left( \frac{2\alpha (t)}{\pi } \right)
^{3/4} \exp{(-\alpha (t) {\bf X}^2)}
\label{eq:kk}
\end{equation}
Note that there is an additional pure exponential factor $\exp
(i\varphi _{\mu}(t))$, where $\varphi _{\mu }$ is the phase of $\mu
(t)$ ($\mu (t)=|\mu (t)| \exp (i\varphi _{\mu } (t))$), but it is
irrelevant for probabilities and can be included in $f({\bf X},t)$.

The above equation shows that, although we have used an
oscillator-type interaction, the particles can become well
separated. The solution only depends on ${\bf X}$. The variable
${\bf Y}$ is not constrained by Eq. ({\ref{eq:kk}}) and can reach
arbitrarily large values.

Now, we consider the decaying system in free evolution. Its
mathematical form can be derived from the equations in interaction
using the relation
\begin{equation}
\lim _{\omega \rightarrow 0} \frac{\sin \omega t}{\omega } =t
\label{eq:ll}
\end{equation}
First, we note that from the term $\exp(if)$ it is simple to see that
the wavefunction at times $t>0$ has no longer the form of a minimum
uncertainty wavepacket (as it is well-known \cite{Gal}, the same
behavior occurs for one-particle minimum uncertainty wavepackets).
The normalized wavefunction (in the relative sense) is given by Eq.

(\ref{eq:kk}) with $\alpha (t)$ replaced by $\alpha _F (t)$, given by
\begin{equation}
\alpha _F (t)=\frac{1}{a^2+\left( \frac{2\hbar t}{Ma} \right) ^2 }
\end{equation}

\section{Correlations}

We analyze in this section if the introduction of the interaction
improves the behavior of the correlations. Two types of measures for the
correlations will be used. One is the probability of detection with
perfect (classical) correlation, the other is the alignment at large
times. We study them separately.

\subsection{Perfect correlations}

Two particles have perfect (classical) correlations when the
positions of the particles found in a measurement process are
opposite: ${\bf x}_{mea}=-{\bf y}_{mea}$ or ${\bf X}_{mea}={\bf
0}$.

Let us first consider the free evolution case. From the expression
for the relative probability density for ${\bf X}={\bf 0}$ we see
that the probability of perfect correlation goes as $\alpha
_F^{3/2}(t)$, a decreasing function of time. For very large times,
$\alpha _F (t)\rightarrow 0$ and the probability of perfect correlation
becomes negligible.

We consider now the interacting case, where $\alpha (t)$ shows an
oscillatory behavior. It varies between $\alpha _- = 1/a^2$ and
$\alpha _+ =(M\omega a/2\hbar )^2$. Depending on the strength of the
interaction coupling $\omega $ we have three different scenarios:
when $\omega > 2\hbar /Ma^2$ we have $\alpha _+ > \alpha _-$,
whereas for $\omega < 2\hbar /Ma^2$ the relation is $\alpha _+ <
\alpha _-$. In the limiting case $\omega = 2\hbar /Ma^2$ we obtain
$\alpha _+ = \alpha _-$. The probability of perfect correlation goes
as $\alpha ^{3/2}(t)$ and oscillates between $\alpha _-^{3/2}$ and
$\alpha _+^{3/2}$. For instance, for $\omega > 2\hbar /Ma^2$ the
probability of perfect correlation increases in the interval $t
\in (0,\pi /2)$ reaching its maximum value at $t=\pi /2$,
starting a stage of decreasing behavior at this point. The perfect
correlations are not completely lost even in the limit of very large times, as
it was the case for free evolution. The analysis for $\omega <
2\hbar /Ma^2$ follows similar lines. The limiting situation occurs
when $\omega = 2\hbar /Ma^2$. Now $\alpha $ is constant and the
probability of perfect correlation detection does not change with
time. We conclude that the presence of the interaction improves the
behavior of the perfect correlations, which become an oscillating or
constant function instead of the decreasing one associated with
free evolution.

\subsection{Alignment}

Alignment is a weaker measure of correlations than perfect
correlations. We say that two particles are aligned when they are
detected in, approximately, opposite directions. A quantitative
criterion for alignment can be introduced following Ref.
\cite{bel}. The angle characterizing angular deviation (perfect
alignment is given by $\pi $) can be expressed as:
\begin{equation}
\tan \theta =\frac{TD(t)}{R(t)}=\frac{\Delta X_i}{<x_i>}
\end{equation}
where $TD(t)$ is the transversal deviation and $R(t)$ is the
distance both particles have traveled. In statistical terms $TD$ can
be taken of the order $\Delta X_i$, i. e., the variance of any of
the components of the center of mass position ($i=1,2 \, or \, 3$,
the problem is isotropic). On the other hand, $R$ is of the order
$<x_i>$, the expectation value of the position of any of the
particles.

We evaluate now these two variables.  We use the Heisenberg picture,
where the evolution of any operator $A$ (we omit the operator symbol
to simplify the notation) is ruled by the equation $i\hbar dA/dt=[
A,H ] + i \hbar \partial A /\partial t$, with $H$ the Hamiltonian of
the system. For the position and momentum operators these relations
become
\begin{equation}
m\frac{d{\bf x}}{dt}={\bf p_x} \; ; \; m\frac{d{\bf y}}{dt}={\bf p_y} \; ; \;
\frac{d{\bf p_x}}{dt}=-2\kappa ({\bf x}+{\bf y})=\frac{d{\bf p_y}}{dt}
\end{equation}
Using the center of mass coordinate ${\bf X}$ and the total
momentum ${\bf P}={\bf p_x}+{\bf p_y}$, the equations can be
rewritten as
\begin{equation}
m\frac{d{\bf X}}{dt}=\frac{1}{2}{\bf P} \; ; \; \frac{d{\bf
P}}{dt}=-8\kappa {\bf X}
\end{equation}
The solution of these equations is
\begin{equation}
{\bf X}={\bf X_o} \cos \Omega t +\frac{{\bf P_o}}{2m\Omega } \sin
\Omega t
\end{equation}
and
\begin{equation}
{\bf P}=-2m\Omega {\bf X_o} \sin \Omega t +{\bf P_o} \cos \Omega t
\end{equation}
with $\Omega ^2=4\kappa /m$ (note $\omega ^2=2\Omega ^2$) and ${\bf
X_o}$ and ${\bf P_o}$ the initial values of both operators.

Assuming the expectation values of both initial conditions to be
null, $<{\bf X_o}>={\bf 0}$ and $<{\bf P_o}>={\bf 0}$, the
variance of $X_i$ can be expressed in the simple form:

\begin{equation}
(\Delta X_i (t))^2 =(\Delta X_{oi})^2 \cos ^2 \Omega t + (\Delta
P_{oi})^2 \frac{\sin ^2 \Omega t}{4M^2 \Omega ^2} +
<X_{oi}P_{oi}> \frac{ \sin \Omega t \cos \Omega t }{M\Omega }
\end{equation}
Finally, we evaluate $<x_i(t)>$. First, we note that from $dp_{x_i}/dt=-2 \kappa X_i$ we have
\begin{equation}
p_{x_i}(t)= p_{x_i}(0)-\frac{2\kappa X_{oi}}{\Omega} \sin \Omega t +
\frac{\kappa P_{oi}}{m\Omega ^2}( \cos \Omega t -1)
\end{equation}
Taking mean values and remembering our assumption $<{\bf X_o}>={\bf
0}$ and $<{\bf P_o}>={\bf 0}$ we have $<p_{x_i}(t)>= <p_{x_i}(0)>$.
Now, we use the relation $mdx_i/dt=p_{x_i}$ that gives (assuming
$<x_i(o)>=0$), $m<x_i(t)>=<p_{x_i}>t$, or

\begin{equation}
<x_i(t)>= <p_{x_i}(0)>t/m
\label{eq:uu}
\end{equation}
Using again the relation (\ref{eq:ll}), we have for the free evolution
\begin{equation}
(\Delta X_i^F (t))^2 =(\Delta X_{oi})^2 + (\Delta P_{oi})^2 \frac{t
^2 }{4M^2 } +\frac{t}{M } <X_{oi}P_{oi}>
\end{equation}
On the other hand, Eq. (\ref{eq:uu}) is valid for both free and interacting evolutions.

Now, we can evaluate the angular deviation. We shall concentrate
on the large times behavior. For the free evolution of the minimum
uncertainty state we have (the $\infty $ superscript refers to $t \rightarrow \infty $):
\begin{equation}
\tan \theta _F^{\infty } = \frac{m\Delta P_{oi}}{2M<p_{x_i}(0)>}
=\frac{< \lambda >}{16\pi \Delta X_{oi}}
\label{eq:WW}
\end{equation}
where we have used the relation (\ref{eq:ww}) and the equation
$<p_{x_i}(0)>=h/< \lambda >$, with $< \lambda >$ the mean wavelength associated
with the initial mean momentum. Equation (\ref{eq:WW}) shows that at
large times there is only alignment if $< \lambda > \ll \Delta X_{oi}$,
i. e., if the mean initial wavelength of the particles is much smaller
than the dispersion of the center of mass of the source.

On the other hand, when the interaction is present the angular
deviation at large times for states with $<p_{x_i}(0)> \neq 0$ is:
\begin{equation}
\tan \theta ^{\infty } =\lim _{t \rightarrow \infty } \frac{m\Delta
X_i}{<p_{x_i}(0)> t}=0
\label{eq:yy}
\end{equation}
because $\Delta X_i$ is a finite function of time due to its
periodic behavior. With interaction there is alignment at large
times independently of the relation between the mean initial
wavelength and the dispersion of the center of mass of the source.

We remark that result (\ref{eq:WW}) is only valid for minimum
uncertainty states (or those obeying the approximate relation
$2\Delta P_{oi} \Delta X_{oi} \approx \hbar $), whereas Eq.

(\ref{eq:yy}) is valid for a much larger class of states, those with
$<P_{ oi}>= <X_{oi}>=0$ and $<p_{x_i}(0)> \neq 0$.

\section{Entanglement}

In the previous section we have analyzed the correlations existent
in the system. Another way to study the problem is to consider the
entanglement present in the wavefunction. The correlations are the
manifestation of the entanglement in measurement processes. A
measure of the entanglement degree is given by the Schmidt number
\cite{Ebe, Fed1, Fed2}. It has been used in a series of studies with important
resemblances with our work \cite{Fed1,Fed2}. In particular, in
\cite{Fed1} it was studied the entanglement existent between a
photon and the atom that has emitted it.

The Schmidt number \cite{Ebe, Fed1} is given by
\begin{equation}
S=\frac{1}{Tr _{\bf x}(\hat{\rho }_{\bf x}^2)}= \frac{1}{Tr _{\bf y}(\hat{\rho }_{\bf y}^2)}
\end{equation}
with the reduced density matrices
\begin{equation}
\hat{\rho }_{\bf x} = Tr _{\bf y} (|\psi ><\psi |) \; ; \; \hat{\rho }_{\bf y} = Tr _{\bf x} (|\psi ><\psi |)
\end{equation}
where $ Tr _{\bf z}$ denotes trace with respect to the variables
associated with particle ${\bf z}$ and $|\psi >$ is given by Eq.
(\ref{eq:kk}). Note that we have used the notation $S$ for the
Schmidt number instead the usual $K$ one in order to avoid any
confussion with the propagator.

In order to evaluate the above traces we must introduce the Fourier
decomposition of the wavefunction
\begin{eqnarray}
\psi ({\bf x},{\bf y},t)=\frac{1}{(2\pi \hbar)^3} \int \int d^3 {\bf p} d^3 {\bf q} \phi ({\bf p},{\bf q},t)
e^{i{\bf p}\cdot {\bf x}/\hbar} e^{i{\bf q}\cdot {\bf y}/\hbar}= \nonumber \\
\int \int d^3 {\bf p} d^3 {\bf q} \phi ({\bf p},{\bf q},t) <{\bf x}|{\bf p}> <{\bf y}|{\bf q}>
\end{eqnarray}
where we have used the usual expression $<{\bf x}|{\bf p}>=(2\pi \hbar)^{-3/2} \exp(i{\bf p}\cdot {\bf x}/\hbar)$.

A simple calculation gives
\begin{equation}
\phi ({\bf p},{\bf q},t) =\sqrt{8} \left( \frac{\pi }{2\alpha (t)} \right) ^{3/4} \exp \left(  \frac{-{\bf q}^2}{4\hbar ^2 \alpha (t)}  \right) \delta ^3 ({\bf p}-{\bf q})
\end{equation}
On the other hand, we have
\begin{eqnarray}
\hat{\rho} _{\bf x}=Tr_{\bf q}(|\psi ><\psi |)=\int d^3 {\bf q} <{\bf q}|\psi ><\psi |{\bf q}> =\nonumber \\
\int \int \int d^3 {\bf q} d^3 {\bf p} d^3 {\bf p_*} \phi ({\bf p},{\bf q},t) \phi ^* ({\bf p_*},{\bf q},t) <{\bf x}|{\bf p}> <{\bf p_*}|{\bf x}>
\end{eqnarray}
where the trace is over the momentum ${\bf q}$ associated
with the variable ${\bf y}$ and we have used the relations $<{\bf
q}|<{\bf y}|{\bf Q}> <{\bf Q_*}|{\bf y}>|{\bf q}>=<{\bf y}|{\bf
y}>\delta ^3({\bf q}-{\bf Q})\delta ^3 ({\bf Q_*}-{\bf q})$.

The second trace is
\begin{eqnarray}
Tr _{\bf p}(\hat{\rho }_{\bf x}^2) = \int d^3 {\bf p} <{\bf p}|\hat{\rho }_{\bf x}^2 |{\bf p}>= \\
\int \int \int \int d^3 {\bf p} d^3 {\bf q} d^3 {\bf p_*} d^3 {\bf q_*} \phi ({\bf q},{\bf p},t) \phi ^*({\bf q},{\bf q_*},t) \phi ({\bf p_*},{\bf q_*},t) \phi ^*({\bf p_*},{\bf p},t) \nonumber
\end{eqnarray}
where the trace now is over the momentum ${\bf p}$ associated with
the variable ${\bf x}$. Note that in the original approach
\cite{Ebe}, $S$ counted the number of effective modes in the Schmidt
decomposition of $|\psi >$. Here, we do not consider the Schmidt
decomposition of the wavefunction but the decomposition on the plane
wave basis. Now the inverse of $ Tr _{\bf p}(\hat{\rho }_{\bf x}^2)$
counts the number of effective modes in this basis. This number is
still denoted as the Schmidt number (see Ref. \cite{Fed2} where this
terminology is also used although there is not a Schmidt
decomposition of the analyzed system).

A simple calculation gives
\begin{equation}
Tr _{\bf p}(\hat{\rho }_{\bf x}^2) = \sqrt{8} (2\pi \hbar )^3 \left( \frac{\pi}{2\alpha (t)}   \right) ^{3/2}
\end{equation}
This equation shows that $ Tr _{\bf p}(\hat{\rho }_{\bf x}^2)$ has
dimensions of $momentum ^3 longitude ^6$. In order to obtain a
dimensionless Schmidt's number the usual expression above must be
replaced. One dimensional parameter provided by the problem is
$\hbar $, which suggests to use $(2\pi \hbar )^3 /Tr _{\bf
p}(\hat{\rho }_{\bf x}^2)$. This expression still has dimensions of
$longitude ^{-3}$, reflecting the fact that the wavefunction is
normalized as a relative probability density per unit volume. Thus,
associated with this wavefunction we cannot have entanglement but
entanglement per unit volume. Then the entanglement in a region of
volume ${\cal V}$ is measured by the following generalization of
Schmidt's number
\begin{equation}
S_{\cal V}=\frac{(2\pi \hbar )^3{\cal V}}{ Tr _{\bf p}(\hat{\rho }_{\bf x}^2)}
= \frac{{\cal V}}{\sqrt{8}} \left( \frac{2\alpha (t)}{\pi }   \right) ^{3/2}
\end{equation}

Using this equation we obtain
\begin{equation}
S_{\cal V}(t) \sim \alpha (t)^{3/2}
\label{eq:son}
\end{equation}
In particular, in the free case for large times, $t \gg Ma/2\hbar$,
we have
\begin{equation}
(S_{\cal V})_F^{\gg}(t) \sim t^{-3}
\end{equation}
Equation (\ref{eq:son}) shows that in the interacting case
entanglement is an oscillatory function of time. It can be analyzed
following the lines of subsection 6.1. This behaviour of
entanglement explains the persistence of correlations. As a matter
of fact, it shows the same dependence on time ($\alpha (t)^{3/2}$)
that perfect correlations. On the other hand, in the free case, the
entanglement decreases following a law of type $t^{-3}$. After some
time almost all the initial entanglement is lost. We conclude that
the pictures of the system obtained using correlations and
entanglement are similar.

Finally, we want to remark that in Ref. \cite{Fed1} the relation
between entanglement and uncertainty relations was studied. A
similar analysis could be carried along the same lines in our case.
However, it would enlarge too much the paper. We plan to do it in
future work. We only notice that in that reference it was also
signaled how the uncertainty relations adopt different forms
depending on the type of measurement done on the system,
single-particle or coincidence one. In this context it must be
remarked that in this work (in Sect. 2) we have only considered
single-particle measurements, which are usual to the Heisenberg
uncertainty relations. In a scheme with conditional measurements
these relations would be different \cite{Fed1}.

\section{Discussion}

We have analyzed in this paper how the behavior of the fragments
originated by the decaying of a mother particle is modified when
interactions between the daughter particles or external fields
driving the system are taken into account. We have shown that the
initial correlations present in the system (such as measured by
perfect correlations and alignment) are much better conserved  when
the interaction is taken into account. This property, if
corroborated for more realistic interactions could provide us with
sources of particles with good correlations in those cases where the
free evolution degrades the initial correlations. In this context
note that our analysis also provides a criterion (the relation in
Eq. (\ref{eq:WW}) between $< \lambda >$ and $\Delta X_{oi}$) to
determine when alignment is preserved by the free evolution.

Good correlations could be useful for various purposes, for
instance, in interferometric experiments with two-particle massive
systems. One example in foundational issues is Popper's argument
\cite{Pop, bel}, where the actual experiments \cite{kim} have been
carried out with photons instead of massive particles. When the
experiments are restricted to the framework of quantum optics we
lose the possibility of studying the dependence of the system on the
form of the wavefunction, which is specially important for
interferometric experiments.

As signaled before decaying systems are archetypal in physics. In
the quantum realm they are particularly well suited to analyze the
influence of uncertainty and spreading on the evolution of the
system and, in particular, on its correlations. Then the example
presented here has also a pedagogical interest.

\end{document}